\documentclass[12pt,aps,prb,preprint,showpacs,showkeys]{revtex4}  

\usepackage{amsmath}    
\usepackage{amsfonts}   
\usepackage{amssymb}
\usepackage{graphicx}   
\usepackage{subfigure}

\begin{document}
\baselineskip = 7.5mm \topsep=1mm
\begin{center}{\LARGE\bf Effect of biquadratic exchange on phase transitions of a planar classical Heisenberg ferromagnet}
\vspace{5 mm}\\ M.\v{Z}ukovi\v{c}$^{\rm a*}$, T.Idogaki$^{\rm b}$ and K.Takeda$^{\rm a,b}$ \vspace{5
mm}\\
\end{center}
\noindent $^{\rm a}$ Institute of Environmental Systems, Kyushu University
\newline
$^{\rm b}$ Department of Applied Quantum Physics, Kyushu University \vspace{3 mm}
\newline
\noindent {\bf{Abstract.}} Effect of biquadratic exchange on phase transitions of a planar classical
Heisenberg (or XY) ferromagnet on a stacked triangular lattice is investigated by Standard Monte Carlo
and Histogram Monte Carlo simulations in the region of a bilinear to biquadratic exchange interaction
ratio $J_{1}/J_{2} \leq 1$. The biquadratic exchange is found to cause separate second-order phase
transitions in a strong biquadratic exchange limit, followed by simultaneous dipole and quadrupole
ordering, which is of first order for an intermediate range of the exchange ratio and changes to a
second-order one again as $J_{1}/J_{2}$ is further increased. Thus, a phase diagram featuring both
triple and tricritical points is obtained. Furthermore, a finite-size scaling analysis is used to
calculate the critical indices for both dipole and quadrupole kinds of ordering. \vspace{20 mm} \\
$PACS\ codes$: 75.10.Hk; 75.30.Kz; 75.40.Cx; 75.40.Mg.
\newline
$Keywords$: Planar Heisenberg ferromagnet; Biquadratic exchange; Phase transition; Quadrupole ordering;
Multicritical point; Histogram Monte Carlo simulation;\\ \vspace{20 mm} \\ $*$Corresponding author.
\newline
Permanent address: Institute of Environmental Systems, Faculty of Engineering, Kyushu University,
Fukuoka 812-8581, Japan
\\ Tel.: +81-92-642-3811; Fax: +81-92-633-6958
\\ E-mail: milantap@mbox.nc.kyushu-u.ac.jp
\newpage
\noindent {\bf\Large{1.Introduction}} \vspace{3mm}
\newline
\indent It has long been recognized that there are materials in which also biquadratic (or generally
multipolar) interactions may play an important role as in some cases they are comparable or even much
larger than the bilinear ones. The problem of higher-order interactions in systems with Heisenberg
symmetry has been tackled in several mean field approximation (MFA) studies
\cite{sivar72}-\cite{chen-levy1}, by high-temperature series expansion (HTSE) calculations
\cite{chen-levy2}, as well as within a framework of some other approximative schemes
\cite{micnas,chaddha99}. Those studies have shown that such interactions can induce various interesting
properties such as tricritical and triple points, quadrupole ordering, separate dipole and quadrupole
phase transitions etc. Much less attention, however, has been paid to this problem on systems with XY
spin symmetry. Chen $et.al$ \cite{chen-etal1,chen-etal2} calculated transition temperatures and the
susceptibility critical indices for a planar Heisenberg ferromagnet with biquadratic exchange on cubic
lattices by the HTSE method. However, they only investigated the region of $J_{1}/J_{2}$ ratio for which
MFA predicts a second-order transition and hence, they failed to provide a complete phase diagram with
all possible phase transitions and their nature for this model. For example, since MFA predicts
first-order transitions for dipole ordering if $J_{1}/J_{2} < 1$, only quadrupole phase transitions,
which are expected to be of second order \cite{carmesin}, were investigated. Here we note that, in spite
of all the previous conclusions based on various approaches, the rigorous proof of the existence of
dipole long-range order (DLRO), corresponding to the ferromagnetic directional arrangement of spins, and
quadrupole long-range order (QLRO), representing an axially ordered state in which spins can point in
either direction along the axis of ordering, at finite temperature on the classical bilinear-biquadratic
exchange model has only recently been provided independently by Tanaka and Idogaki \cite{tanaka}, and
Campbell and Chayes \cite{campbell}.
\newline
\indent We have recently performed MC simulations on the XY model with the bilinear-biquadratic exchange
Hamiltonian on a simple cubic lattice in order to establish phase boundaries and describe critical
behaviour of the system \cite{nagata}. Unfortunately, we did not succeed in finding a conclusive answer
to the problem of the order of the DLRO transition in the region of the separate DLRO and QLRO
transitions. In the present paper we chose the lattice with a hexagonal (stacked triangular) symmetry,
increased the lattice size and performed a careful finite-size scaling (FSS) analysis in order to obtain
more reliable data and deliver a complete picture of a long-range ordering and its nature for the
considered system via Standard Monte Carlo (SMC) and Histogram Monte Carlo (HMC) simulations. The phase
diagram obtained captures all important features induced by the biquadratic exchange such as separate
ordering of dipoles and quadrupoles, the appearance of triple and tricritical points, short-range order
above the quadrupole ordering temperature, etc. Furthermore, we perform a FSS analysis to determine the
order of a transition from the energy cumulant scaling, and to calculate the critical indices for both
DLRO and QLRO transitions. \vspace{6mm}
\newline
\noindent {\bf\Large{2.Model and simulation technique}}\vspace{3mm}
\newline
\indent We are concerned with the planar classical Heisenberg model with bilinear and biquadratic
interactions, described by the Hamiltonian
\begin{equation}
H=-J_{1}\sum_{\langle i,j \rangle}\mbox{\boldmath $S$}_{i} \cdot \mbox{\boldmath $S$}_{j}
-J_{2}\sum_{\langle i,j \rangle}(\mbox{\boldmath $S$}_{i} \cdot \mbox{\boldmath $S$}_{j})^{2} \ ,
\end{equation}
where $\mbox{\boldmath $S$}_{i} = (S_{ix},S_{iy})$ is a two-dimensional unit vector at the $i$th lattice
site , $\langle i,j \rangle$ denotes the sum over nearest neighbors, and $J_{1},J_{2}>0$ are the
bilinear and biquadratic exchange interaction constants, respectively.
\newline
\indent We first perform SMC simulations on systems of $N=L^{3}$ spins, where $L$ = 12, 15, 18, 24, 30,
assuming periodic boundary condition throughout. For a fixed value of the exchange ratio $J_{1}/J_{2}$
we run simulations starting at low (high) temperatures from a ferromagnetic (random) initial
configuration and gradually raise (lower) temperature. These heating-cooling loops serve to check
possible hysteresis, accompanying first-order transitions. As we move in $(J_{1}/J_{2},k_{B}T/J_{2})$
space, we use the last spin configuration as an input for calculation at the next point. We sweep
through the spins in sequence and updating follows a Metropolis dynamics. In the updating process, the
new direction of spin in the spin flip is selected completely at random, without any limitations by a
maximum angle of spin rotation or allowed discrete set of resulting angle values. Thermal averages are
calculated using at most $1\times10^{5}$ Monte Carlo steps per spin (MCS/s) after equilibrating over
another $0.5\times10^{5}$ MCS/s. We calculate the system internal energy $E$ and some other physical
quantities defined as follows:
\newline
the specific heat per site $c$
\begin{equation}
\label{eq.c}c=\frac{(\langle E^{2} \rangle - \langle E \rangle^{2})}{Nk_{B}T^{2}}\ ,
\end{equation}
the DLRO parameter (magnetization) $m$,
\begin{equation}
\label{eq.m}m=\langle M \rangle/N,\ {\mathrm{where}}\
M=\left[\left(\sum_{i}S_{ix}\right)^{2}+\left(\sum_{i}S_{iy}\right)^{2}\right]^{\frac{1}{2}}\ ,
\end{equation}
the QLRO parameter $q$,
\begin{equation}
\label{eq.q}q=\langle Q \rangle/N,\ {\mathrm{where}}\
Q=\left[\left(\sum_{i}\left(\left(S_{ix}\right)^{2}-\left(S_{iy}\right)^{2}\right)\right)^{2}+
\left(\sum_{i}2S_{ix}S_{iy}\right)^{2}\right]^{\frac{1}{2}}\ ,
\end{equation}
and the following quantities which are functions of the parameter $O$ ($=\ M,\ Q$) :
\newline
the susceptibility per site $\chi_{O}$
\begin{equation}
\label{eq.chi}\chi_{O} = \frac{(\langle O^{2} \rangle - \langle O \rangle^{2})}{Nk_{B}T}\ ,
\end{equation}
the logarithmic derivatives of $\langle O \rangle$ and $\langle O^{2} \rangle$ with respect to
$K=1/k_{B}T$
\begin{equation}
\label{eq.D1}D_{1O} = \frac{\partial}{\partial K}\ln\langle O \rangle = \frac{\langle OE
\rangle}{\langle O \rangle}- \langle E \rangle\ ,
\end{equation}
\begin{equation}
\label{eq.D2}D_{2O} = \frac{\partial}{\partial K}\ln\langle O^{2} \rangle = \frac{\langle O^{2} E
\rangle}{\langle O^{2} \rangle}- \langle E \rangle\ ,
\end{equation}
the fourth-order long-range order (LRO) cumulants $U_{1}$ and $U_{2}$
\begin{equation}
\label{eq.U1U2}U_{1} = 1-\frac{\langle O^{4}\rangle}{3\langle O^{2}\rangle^{2}} \ , \ \
U_{2}=2-\frac{\langle O^{4}\rangle}{\langle O^{2}\rangle^{2}}\ ,
\end{equation}
and the fourth-order energy cumulant $V$
\begin{equation}
\label{eq.V}V = 1-\frac{\langle E^{4}\rangle}{3\langle E^{2}\rangle^{2}}\ .
\end{equation}
These quantities are used to estimate the nature of a transition as well as its position. First-order
transitions usually manifest themselves by discontinuities in the order parameter and energy, and
hysteresis when cooling and heating. If transition is second order, it can be approximately localized by
the $\chi_{O}$ peak position and also double-checked by the position of the fourth-order LRO and energy
cumulants curves intersection for various lattice sizes.
\newline
\indent In the next stage we perform HMC calculations, developed by Ferrenberg and Swendsen
\cite{ferr-swen1,ferr-swen2}, at the transition temperatures estimated from the SMC calculations for
each lattice size. Here, $2\times10^{6}$ MCS are used for calculating averages after discarding another
$1\times10^{6}$ MCS for thermalization. We calculate the energy histogram $P(E)$, the order parameters
histograms $P(O)$\ $(O=M,Q)$, as well as the physical quantities (\ref{eq.c})-(\ref{eq.V}). Using data
from the histograms one can calculate physical quantities at neighboring temperatures, and thus
determine the values of extrema of various quantities and their locations with high precision for each
lattice size. In such a way we can obtain quality data for FSS analysis which determines the order of
the transition and, in the case of a second-order transition, it also allows us to extract critical
indices. For example, the energy cumulant $V$ exhibits a minimum near critical temperature $T_{c}$,
which achieves the value $V^{*}=\frac{2}{3}$ in the limit $L\rightarrow \infty$ for a second-order
transition, while $V^{*}<\frac{2}{3}$ is expected for a first-order transition
\cite{ferr-swen1,ferr-swen2}. The extrema of a variety of thermodynamic quantities at a second-order
transition are known to scale with a lattice size as, for example:
\begin{equation}
\label{eq.scalchi}\chi_{O,max}(L) \propto L^{\gamma_{O}/\nu_{O}}\ ,
\end{equation}
\begin{equation}
\label{eq.scalV1}D_{1O,max}(L) \propto L^{1/\nu_{O}}\ ,
\end{equation}
\begin{equation}
\label{eq.scalV2}D_{2O,max}(L) \propto L^{1/\nu_{O}}\ ,
\end{equation}
\noindent where $\nu_{O}$ and $\gamma_{O}$ represent the correlation length and susceptibility critical
exponents, respectively. In the case of a first-order transition (except for the order parameters), they
display a volume-dependent scaling, $\propto L^{3}$. The simulations were performed on the vector
supercomputer FUJITSU VPP700/56.\vspace{6mm}
\newline
\noindent {\bf\Large{3.FSS analysis and phase diagram}}\vspace{3mm}
\newline
\indent We started our calculations in a region where the biquadratic exchange is equal to the bilinear
one, i.e. $J_{1}/J_{2}=1$, and gradually lowered the ratio down to 0. For the ratio greater than
$J_{1}/J_{2} \simeq 0.55$ we found only one phase transition corresponding to a dipole long-range
ordering which is of second order. Hence, as far as phase transitions are concerned, in this range of
the exchange ratio the presence of the biquadratic exchange is only reflected in the position of the
transition temperature (as well as critical indices to some extent) but causes no qualitative changes
compared with the case of $J_{2}=0$ \cite{guillou,janke}. The second-order character of a transition in
this region was also confirmed by HMC calculations. Neither double-peaked energy or order parameter
histogram nor volume-dependent scaling of the calculated quantities were observed. Fig.1 shows the FSS
calculation of $\nu_{M}$ from Eqs. (\ref{eq.scalV1}) and (\ref{eq.scalV2}), and $\gamma_{M}$ from Eqn.
(\ref{eq.scalchi}) in a ln-ln plot, for $J_{1}/J_{2} = 0.8$. The best fit value $\nu_{M}=0.677 \pm
0.008$ \footnote{The errors for $\nu_{O}$ and $\gamma_{O}$ are calculated from standard errors of the
respective slopes $b$ in the linear regression $y=a+bx$.} agrees very well with the values obtained from
other calculations for $J_{2}=0$, but $\gamma_{M}=1.177 \pm 0.015$ is apparently lower than the values
obtained for $J_{2}=0$ ($\nu_{M}=0.669$ and $\gamma_{M}=1.316$ from Refs. \cite{guillou,janke}).
However, the critical indices are expected to change by the presence of the biquadratic exchange
\cite{allan} and, as shown by the series-expansion calculations for $J_{1}/J_{2} \geq 1$
\cite{chen-etal1}, $\gamma_{M}$ is continuously lowered by increasing biquadratic exchange.
\newline
\indent As the exchange ratio is lowered slightly below $J_{1}/J_{2} \simeq 0.55$, the order of the
transition changes. Still no separate quadrupole ordering is observed, however, the transition becomes
apparently first order. This is clearly seen in Fig.2, where bimodal nature of energy distribution
histograms for various lattice sizes at $J_{1}/J_{2}=0.5$ is obvious. With increasing lattice size the
barrier between the two energy states is increasing, indicating that the energy is discontinuous and
hence, the transition of first order. However, as can be seen from Fig.3, for the case of
$J_{1}/J_{2}=0.35$ the bimodal distribution can only be observed at sufficiently large $L$, indicating
vicinity of the multicritical point. We note that similar behaviour was also observed in both order
parameters distribution histograms, although we do not show it here.
\newline
\indent The bimodal distribution vanishes as the ratio drops below the value $J_{1}/J_{2} \simeq 0.33$.
Moreover, below $J_{1}/J_{2} \simeq 0.33$ quadrupoles start ordering separately at temperatures higher
than those for dipole ordering. Thus the phase boundary branches and a new middle phase of axial
quadrupole long-range order (QLRO) without magnetic dipole ordering opens between the paramagnetic and
DLRO phases. This phase broadens as $J_{1}/J_{2}$ decreases, since the QLRO branch is little sensitive
to the $J_{1}/J_{2}$ ratio variation and levels off, while the DLRO branch turns down approaching the
point ($J_{1}/J_{2},k_{B}T/J_{2})=(0,0)$. This means that the ground state is always magnetic as long as
there is a finite dipole exchange interaction. In Fig.4 we present the temperature variation of both
DLRO and QLRO parameters $m$ and $q$, respectively, at $J_{1}/J_{2}=0.1$. We can see that quadrupoles
order before dipoles, forming a fairly broad region of QLRO without DLRO. The question is of what order
are these transitions. MFA predicts a first-order transition to DLRO phase in this region
\cite{chen-levy2,chen-etal1}, although some other MFA results allow possibility of a second-order DLRO
transition in the high biquadratic exchange limit \cite{sivar72}. Unfortunately, no other calculations
by more reliable techniques are available. In fact, determination of the order of a transition here
turned out to be quite tricky. Although no discontinuities in the internal energy temperature variation
were observed, extremely sharp magnetization slopes and consequently, extremely sharp susceptibility
peaks at the transition (Fig.5) would rather suggest a first-order transition. Moreover, the
fourth-order cumulant $U_{2}$ at the transition falls to negative values displaying a minimum, i.e.
displays behaviour seemingly typical for a first-order transition. This, however, must be taken with
precaution since, here, with increasing temperature the system does not enter a paramagnetic phase but a
different ordered phase. Indeed, further investigation showed that the minimum did not scale with the
lattice size as it should in the case of a first-order transition \cite{vollmayr} and, hence, that in
this case it is of different origin and can not be seen as an indicator of a first-order transition.
Therefore, in order to resolve the ambiguities in the phenomena observed from SMC calculations we
performed a finite-size scaling analysis from HMC data. As mentioned in Introduction, a first-order
transition can be reliably detected via the fourth-order energy cumulant $V$ scaling, which is shown in
Fig.6 for $J_{1}/J_{2}=0.1$. As we can see, a minimum near $T_{c}$ extrapolated to the limit
$L\rightarrow \infty$, $V^{*}$, reaches the value very close to $\frac{2}{3}$, as it should be in the
case of a second-order transition. Also the other quantities clearly did not scale with volume. Instead,
as shown in Fig.7, the obtained scaling indices $\nu_{M}=0.642 \pm 0.006$ and $\gamma_{M}=1.241 \pm
0.011$ match quite well the values expected at a second-order transition for the given quantities and
universality class (See comments on the universality class in question in the Discussion and concluding
remarks section). Although the transition at $J_{1}/J_{2}=0.1$ appears to be undoubtedly of second
order, it might change to a first-order one at higher ratio $J_{1}/J_{2}$ but before the DLRO and QLRO
boundaries merge, as suggested in Ref. \cite{sivar72}. Therefore, we checked for such a possibility but
did not find it to be so. Scaling analysis, similar to that for the case of $J_{1}/J_{2}=0.1$, performed
for the ratio $J_{1}/J_{2}=0.3$, which is close to the limiting value $J_{1}/J_{2} \simeq 0.33$,
produced $V^{*}=0.6666(4)$, a value which deviates slightly more from $\frac{2}{3}$ than in the previous
case, but is still too close to be considered as a sign of a first-order transition. Also the critical
indices $\nu_{M} = 0.654 \pm 0.009$ and $\gamma_{M} = 1.216 \pm 0.006$ confirm the second-order nature
of the transition. Therefore, we conclude that the DLRO transition remains second order in the whole
region up to $J_{1}/J_{2} \simeq 0.33$.
\newline
\indent On the other hand, the nature of the separate QLRO transition is much more obvious. All
calculated observables behave typically like those for a second-order transition and so does the
scaling. This is in agreement with the work of Carmesin \cite{carmesin}, who performed mapping of
quadrupolar interactions to dipolar ones for planar systems and, hence, showed that a transition in
these systems should be second order. Critical indices describing a QLRO transition at $J_{1}/J_{2} =
0.2$ are extracted from the scaling depicted in Fig.8. Their magnitudes $\nu_{Q} = 0.660 \pm 0.008$ and
$\gamma_{Q} = 1.305 \pm 0.011$ are in a good agreement with the values obtained for critical indices
connected with a DLRO transition. These values are also in a good agreement with the values obtained
from the high-temperature series expansion calculations for cubic lattices \cite{chen-etal2,nagata}.
\newline
\indent Observing the specific heat temperature dependence we noticed a shoulder-like broadening just
above the quadrupole-ordering transition temperature (Fig.9). We ascribe this phenomenon to short-range
order, which was already predicted to exist by Chen and Levy \cite{chen-levy1}, however, apparently
could not be seen within their MFA scheme. Here, we show a snapshot of the spin arrangement in a XY
plane taken at $T_{1}>T_{Q}$ (inset). In the snapshot we can see that the quadrupoles are not totally
disordered but tend to align locally in regions equally partitioned among the crystalographically
equivalent three directions. At the quadrupole-ordering temperature the quadrupoles choose one ordering
direction and the entire crystal forms a single domain.
\newline
\indent The resulting phase diagram is drawn in Fig.10 and critical indices calculated for second-order
transitions are summarized in Table \ref{tab.1}. \vspace{6mm}
\newline
\noindent {\bf\Large{5.Discussion and conclusions}}\vspace{3mm}
\newline
\indent In this work we studied effects of the biquadratic exchange on the phase diagram of the
classical XY ferromagnet on a stacked triangular lattice (STL). We believe that our investigations,
which are the first of this kind for a system with hexagonal lattice symmetry, covered all significant
phenomena induced by the presence of the biquadratic exchange and present a fairly compact picture of
the role of this higher-order exchange interaction on the critical behaviour of the system considered.
We obtained the phase diagram with two ordered phases, featuring some interesting phenomena such as the
appearance of tricritical and triple points. We found that in the region where the bilinear exchange is
dominant there is only one phase transition to the DLRO phase, which remains second order until the
exchange ratio reaches the value $J_{1}/J_{2} \simeq 0.55$. Upon further lowering of the ratio, the
transition changes to a first-order one at the tricritical point and remains so down to $J_{1}/J_{2}
\simeq 0.33$. Below this value the phase boundary splits into the QLRO transition line at higher
temperatures and the DLRO transition line at lower temperatures, which are both of second order. Hence,
there is a triple point at $J_{1}/J_{2} \simeq 0.33$, where two second-order and one first-order
transition boundaries meet.
\newline
\indent We speculate that the mechanism driving the system through the mentioned transitions could be as
follows. At $J_{1}/J_{2} \rightarrow \infty$ ($J_{2}=0$) the transition is known to be second order. If
we introduce the biquadratic exchange and reduce $J_{1}/J_{2}$, i.e. reduce the bilinear couplings, this
might induce in a sense a kind of tension (competition) between the two exchange interactions. Namely,
while the decreasing bilinear exchange drives the transition temperature down to the lower values, the
biquadratic exchange does not follow this tendency and rather prevents the ordering temperature from
rapid decrease. This tendency is clearly seen from the phase diagram both in the region of separate
transitions, where $T_{Q}$ does not vary much with decreasing $J_{1}/J_{2}$, as well as in the region of
simultaneous ordering, where the transition temperature is apparently enhanced by the presence of the
biquadratic exchange (the case of absent biquadratic exchange is represented by the dash-dot straight
line in ($J_{1}-k_{B}T_{c}$) parameter space). Put differently, quadrupoles would prefer ordering at
higher temperatures but as long as there is a single transition they are prevented to do so by too low
bilinear exchange, and order occurs only if the temperature is lowered still further. This frustration
is enhanced as $J_{1}/J_{2}$ decreases, and below $J_{1}/J_{2} \simeq 0.55$ it results in a first-order
transition when the strength of the quadrupole ordering prevails and frustrated quadrupoles order
abruptly along with dipoles. However, below $J_{1}/J_{2} \simeq 0.33$ the frustration is too high for
the two kinds of ordering to occur simultaneously - they separate, the tension is released and the
transitions become second order again.
\newline
\indent The calculated critical indices in the region $0 \leq J_{1}/J_{2} \leq 0.33$ for QLRO and $0.55
\leq J_{1}/J_{2}$ for DLRO transitions were found to belong to the standard three-dimensional XY
universality class, although they showed a slight variation with changing $J_{1}/J_{2}$, as had also
been observed in the HTSE calculations \cite{chen-etal1,chen-etal2}. The values of the indices for the
DLRO transition in the region $0 < J_{1}/J_{2} \leq 0.33$ appear to be in a reasonable agreement with
those just mentioned above, and one might tend to include them into the same XY universality class. We
believe, however, that they rather belong to the Ising universality class (Note that the critical
indices of both universality classes take fairly similar values: $\nu^{XY} = 0.669$, $\gamma^{XY} =
1.316$ and $\nu^{Ising} = 0.629$, $\gamma^{Ising} = 1.239$ \cite{ferr-landau} for $J_{2}=0$). The reason
for our claim is that in this region there is first axial quadrupole ordering and only then directional
dipole ordering within the given axis. Therefore, the only difference from the Ising case is that
dipoles can order along any of the three axes, not only the z-axis. The Ising-like nature of the
ordering would also explain the remarkably sharp behaviour of the physical quantities at the transition
as, for example, shown in Fig.5.
\newline
\indent We further intend to perform similar simulations on STL with antiferromagnetic bilinear and
antiferromagnetic/ferromagnetic biquadratic exchange. Such a spin system would present a geometrically
frustrated system with additional frustration arising from competition between the antiferromagnetic
bilinear and ferromagnetic biquadratic, or inter-plane antiferromagnetic bilinear and antiferromagnetic
biquadratic exchange interactions. Among the main points of interest of such investigations will be the
nature of the ground-state order, the universality class of the antiquadrupolar ordering, etc.

\newpage

\begin{figure}[!t]
\includegraphics[scale=0.5]{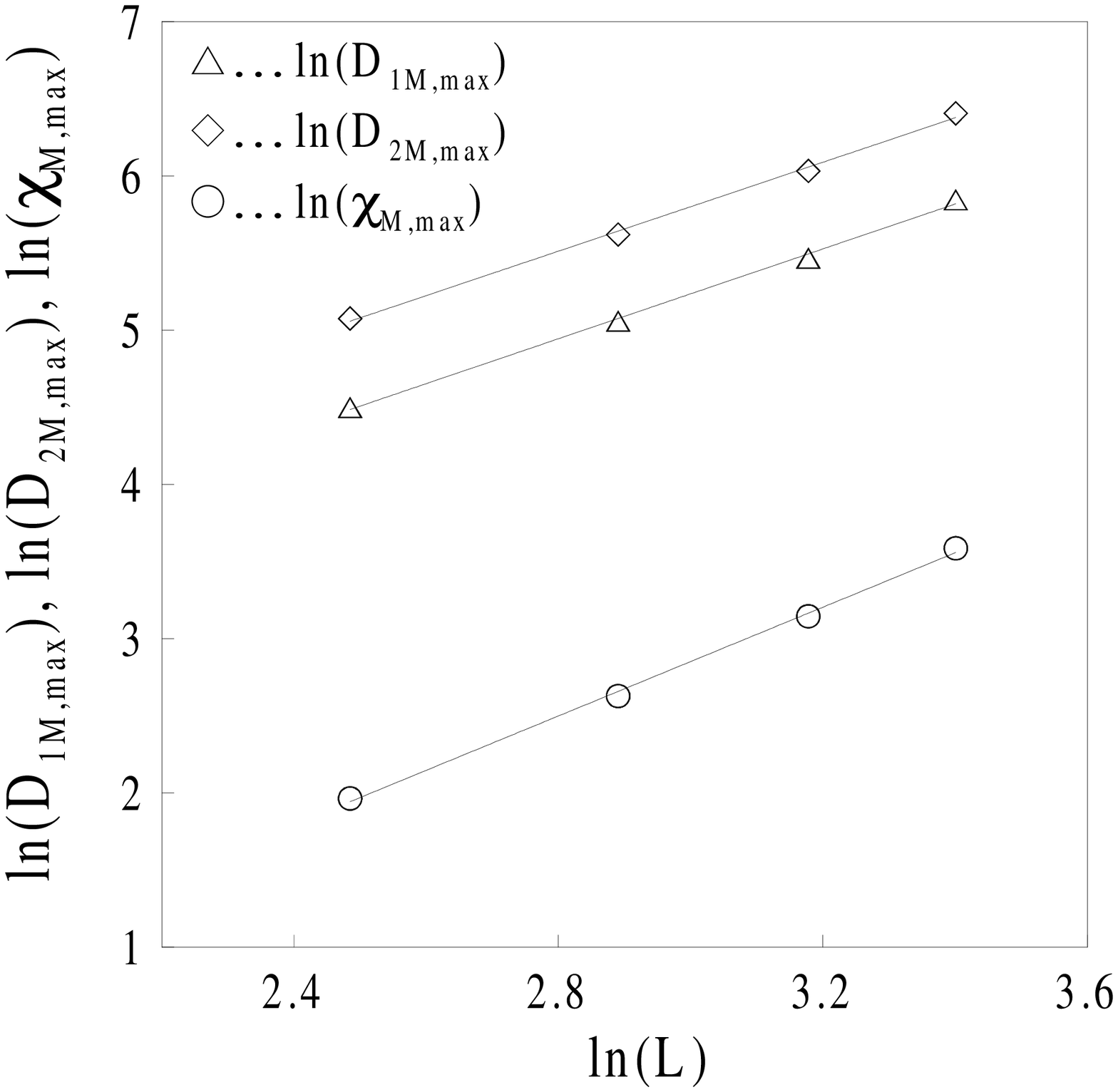}
\caption{Scaling behaviour of the maxima of the susceptibility $\chi_{M,max}$ and
logarithmic derivatives of the DLRO parameter and its second moment $D_{1M,max}$ and $D_{2M,max}$,
respectively, in ln-ln plot, for $J_{1}/J_{2} = 0.8$. The slopes yield values of $1/\nu_{M}$ for
$D_{1M,max},\ D_{2M,max}$ and $\gamma_{M}/\nu_{M}$ for $\chi_{M,max}$.}
\end{figure}

\begin{figure}[!t]
\includegraphics[scale=0.5]{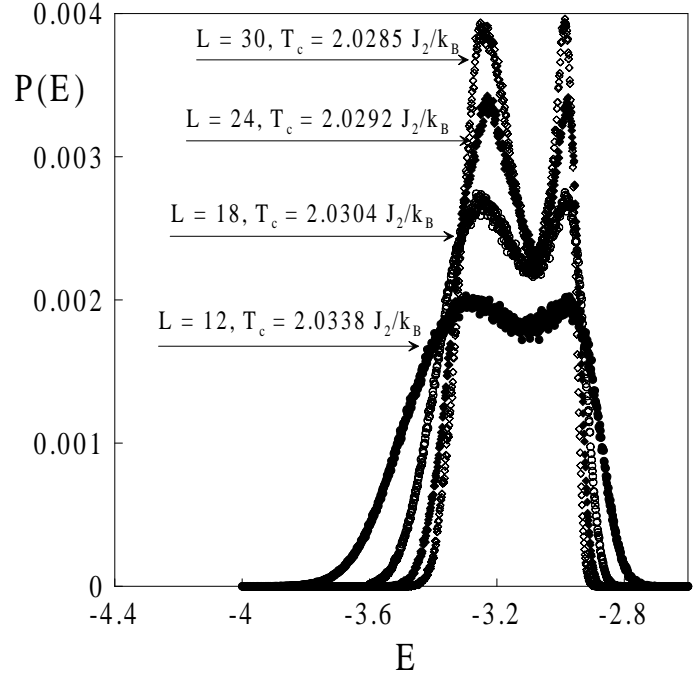}
\caption{Energy distribution at the size-dependent transition temperatures
$T_{c}(L)$ for various lattice sizes and $J_{1}/J_{2} = 0.5$. Double-peaked structure with deepening
barrier between the two energy states with increasing lattice size indicates a first-order transition.}
\end{figure}

\begin{figure}[!t]
\includegraphics[scale=0.5]{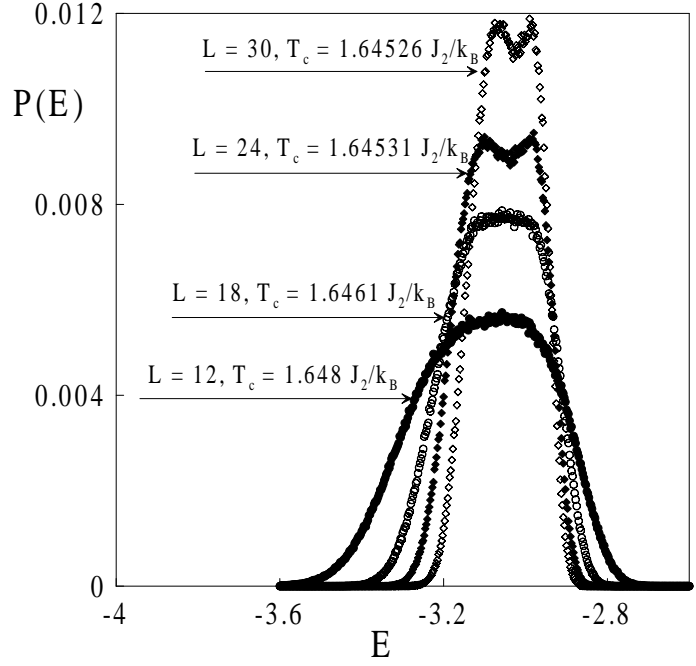}
\caption{Energy distribution at the size-dependent transition temperatures $T_{c}(L)$ for
various lattice sizes and $J_{1}/J_{2} = 0.35$. The bimodal distribution signaling a first-order
transition can only be seen at sufficiently large $L$.}
\end{figure}

\begin{figure}[!t]
\includegraphics[scale=0.5]{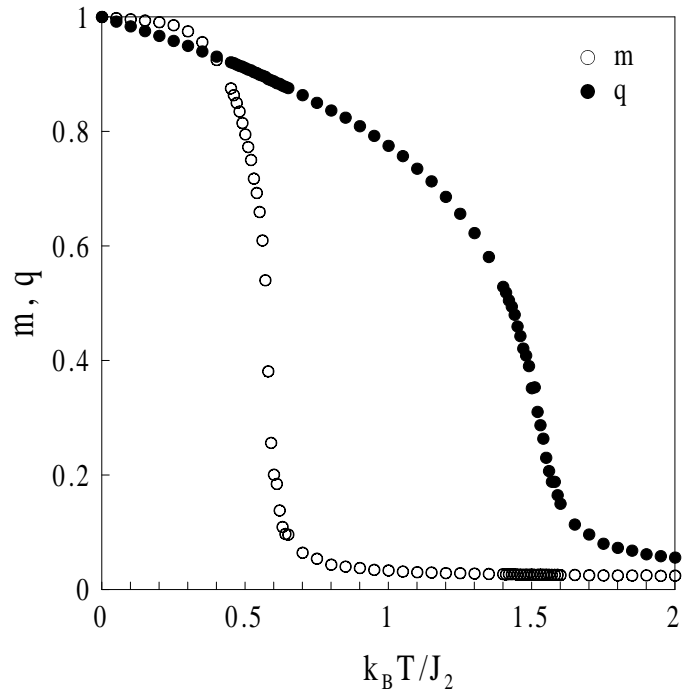}
\caption{Temperature variation of the DLRO and QLRO parameters $m$ and $q$, respectively,
for $J_{1}/J_{2} = 0.1$ and L = 12.}
\end{figure}

\begin{figure}[!t]
\includegraphics[scale=0.5]{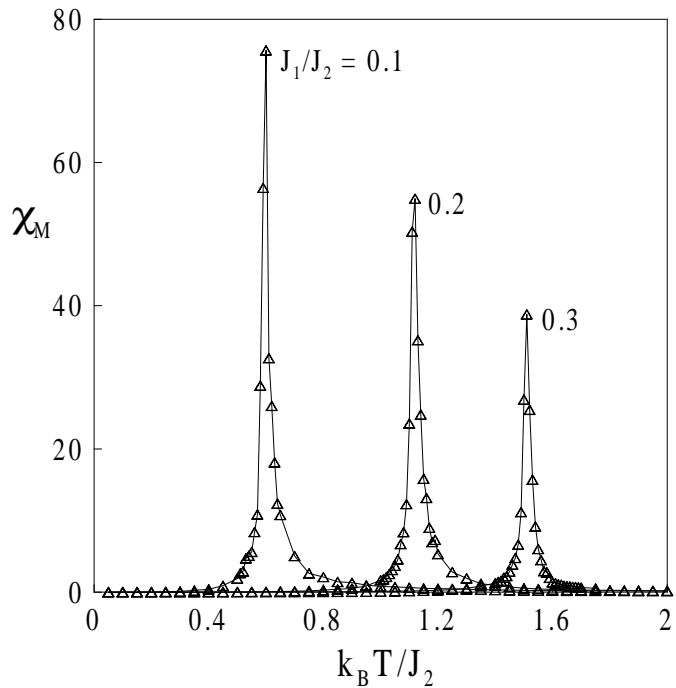}
\caption{Susceptibility $\chi_{M}$ vs temperature for different values of $J_{1}/J_{2}$
and L = 18.}
\end{figure}

\begin{figure}[!t]
\includegraphics[scale=0.5]{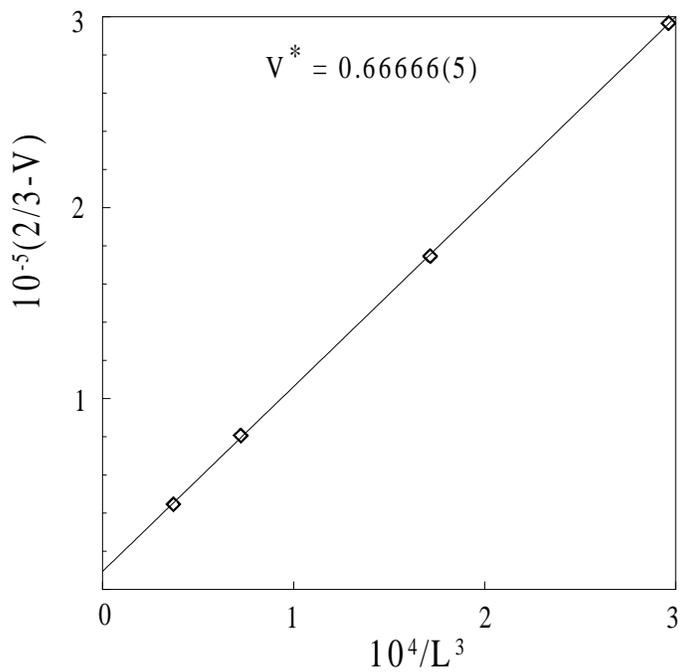}
\caption{Scaling of the energy cumulant minima with volume at $J_{1}/J_{2} = 0.1$. The
value $V^{*}=0.66666(5)$ is obtained from the extrapolation $L \rightarrow \infty$. We note that the
brackets here only signify that the last digit could not be determined with certainty from the best fit
and, hence, the value in the brackets should not be seen as an error estimate.}
\end{figure}

\begin{figure}[!t]
\includegraphics[scale=0.5]{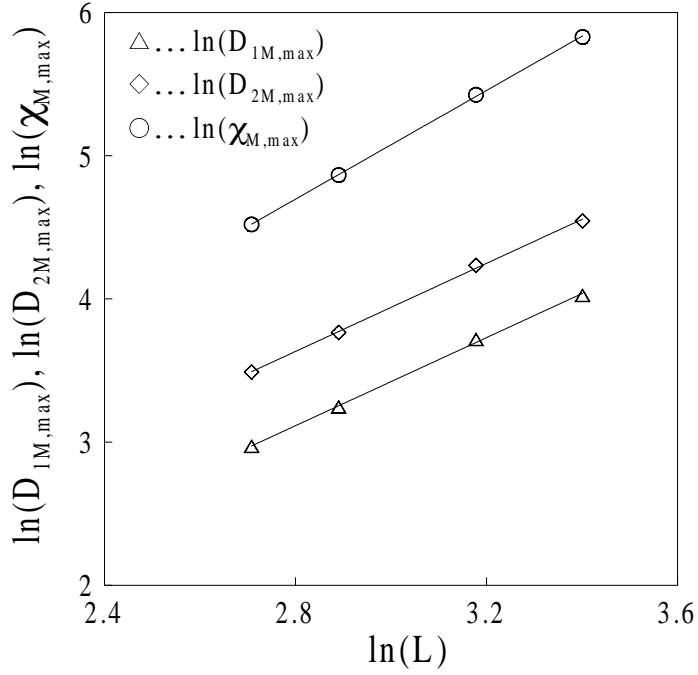}
\caption{Scaling analysis of $\chi_{M,max}$, $D_{1M,max}$ and $D_{2M,max}$ as in Fig.1,
for the case of $J_{1}/J_{2} = 0.1$.}
\end{figure}

\begin{figure}[!t]
\includegraphics[scale=0.5]{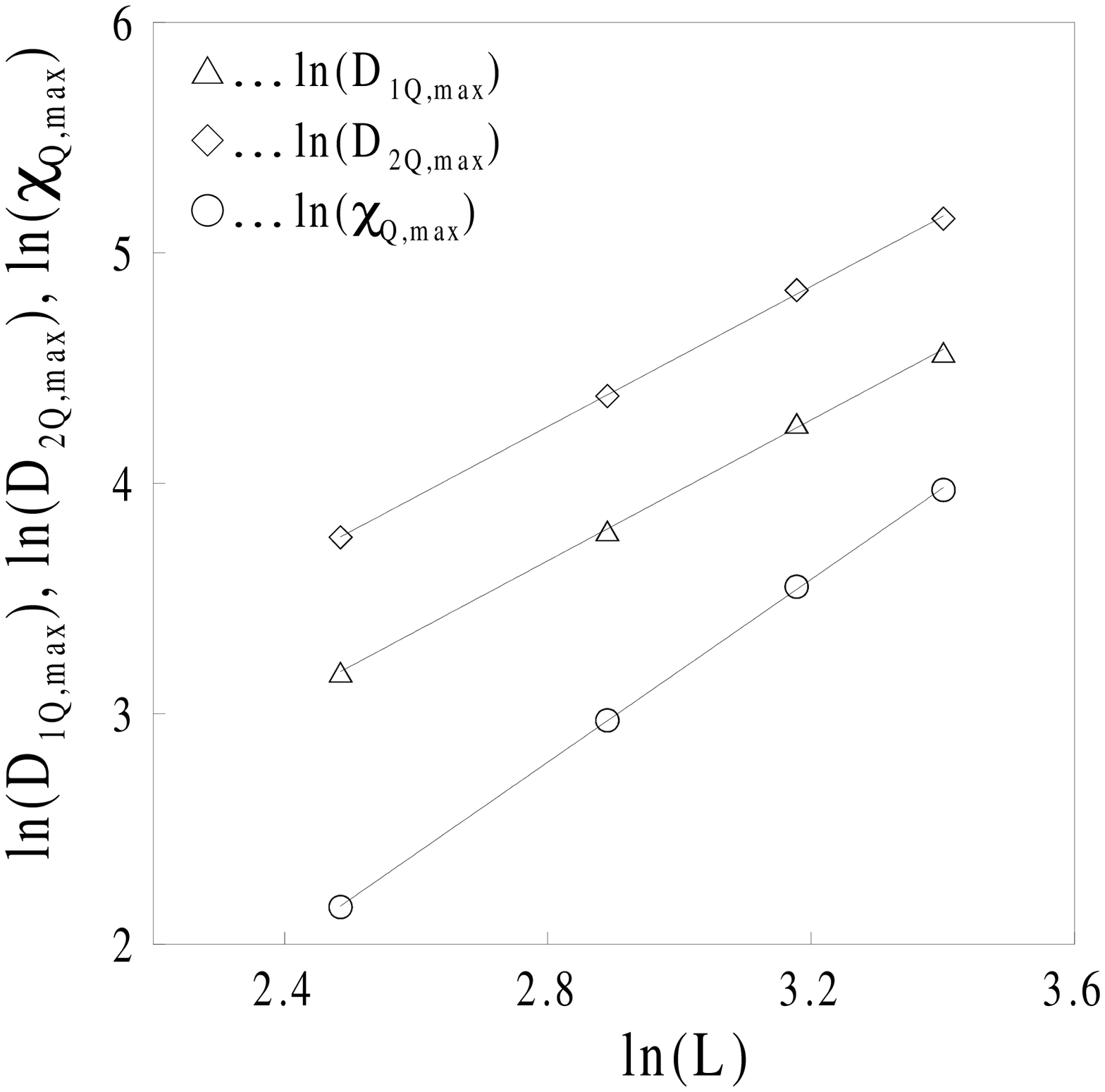}
\caption{Scaling behaviour of the maxima of the susceptibility $\chi_{Q,max}$ and
logarithmic derivatives of the QLRO parameter and its second moment $D_{1Q,max}$ and $D_{2Q,max}$,
respectively, in ln-ln plot, for $J_{1}/J_{2} = 0.2$. The slopes yield values of $1/\nu_{Q}$ for
$D_{1Q,max},\ D_{2Q,max}$ and $\gamma_{Q}/\nu_{Q}$ for $\chi_{Q,max}$.}
\end{figure}

\begin{figure}[!t]
\subfigure{\includegraphics[scale=0.5]{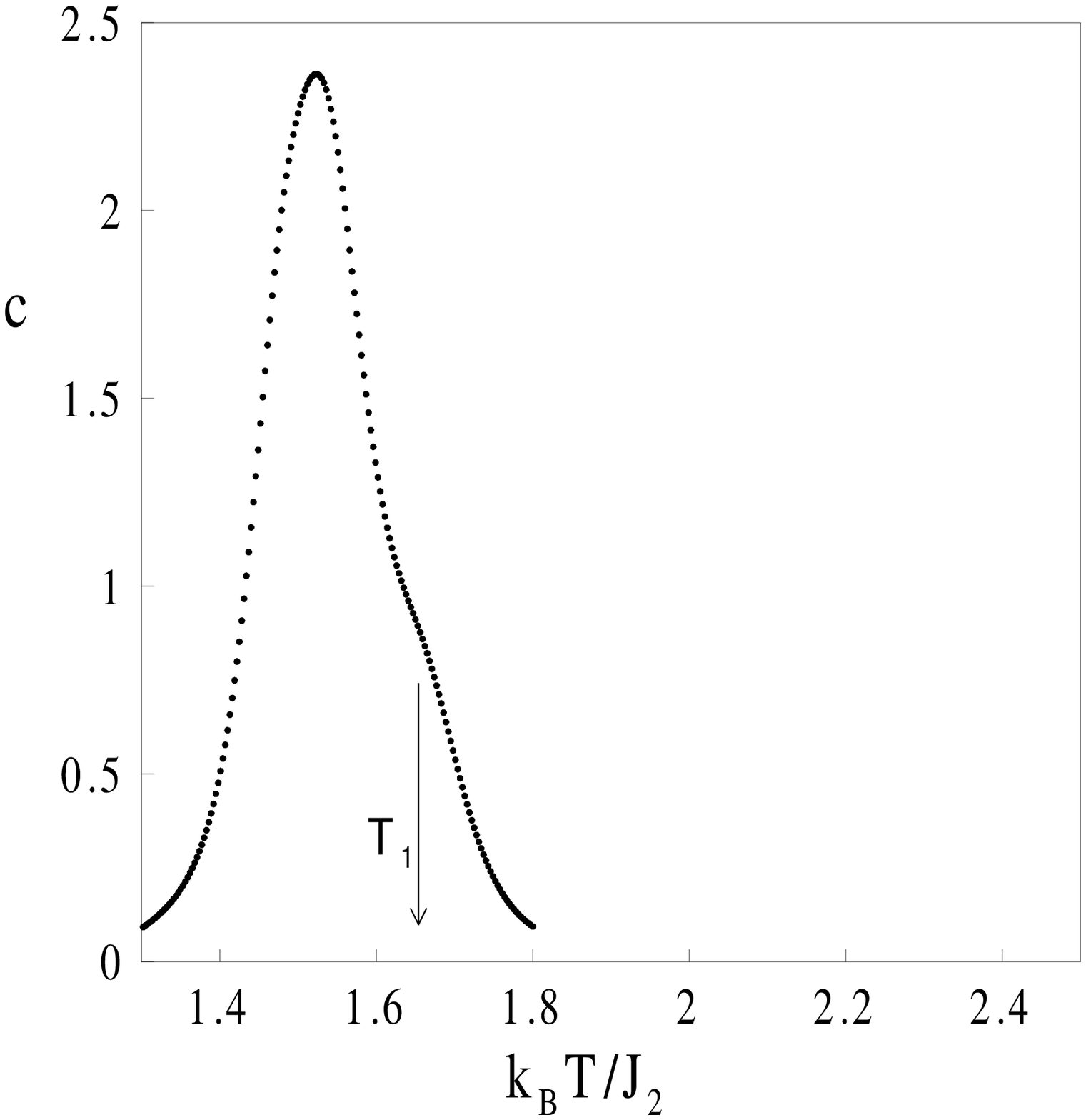}}
\subfigure{\includegraphics[scale=0.5]{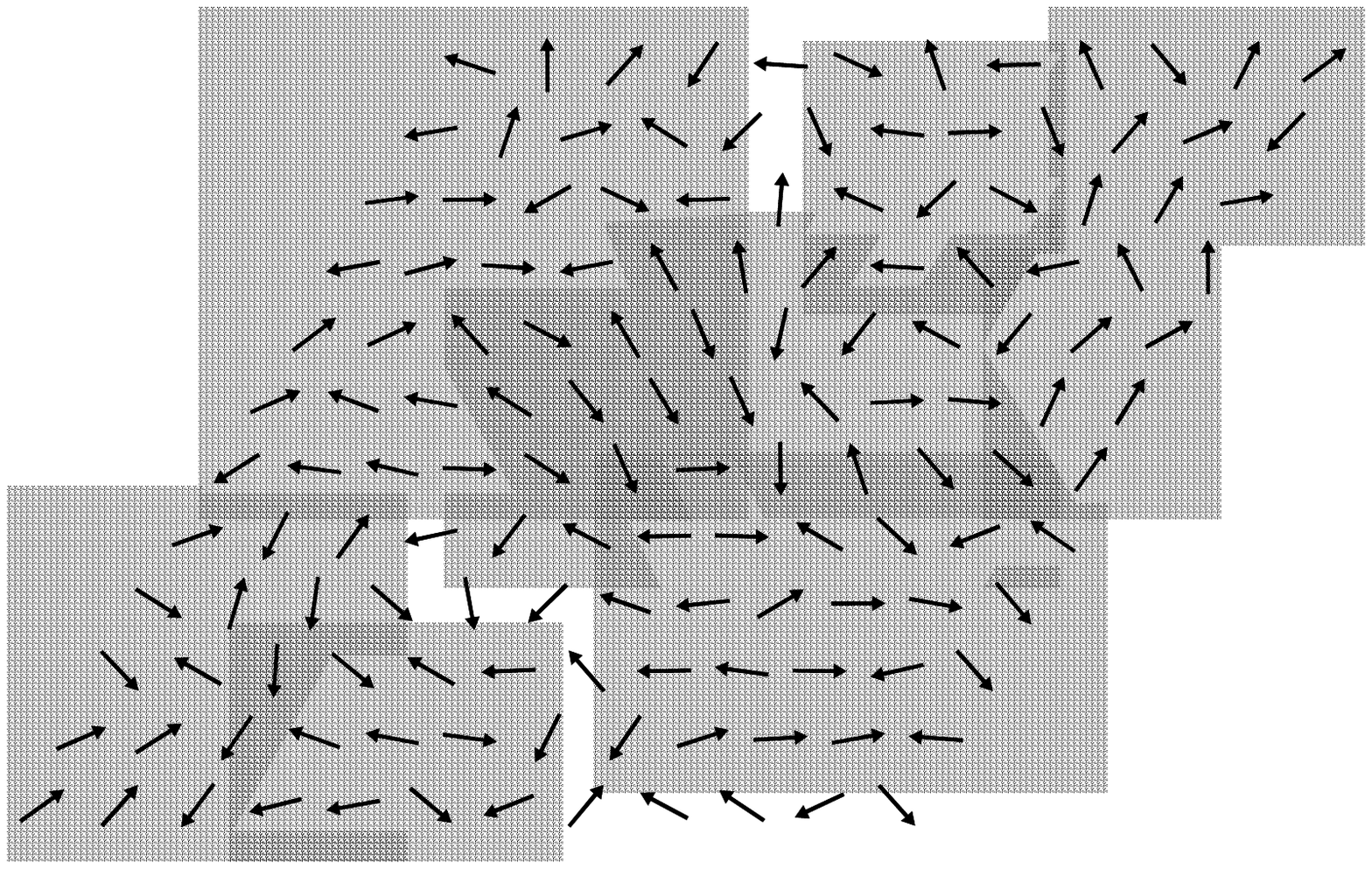}}
\caption{(a) Specific heat vs temperature showing a shoulder-like broadening at $T_{1}>T_{Q}$,
indicating the presence of short-range order above a quadrupole ordering temperature. (b) Snapshot shows signs 
of a local alignment of spins along the crystalographically equivalent three directions.}
\end{figure}

\begin{figure}[!t]
\includegraphics[scale=0.5]{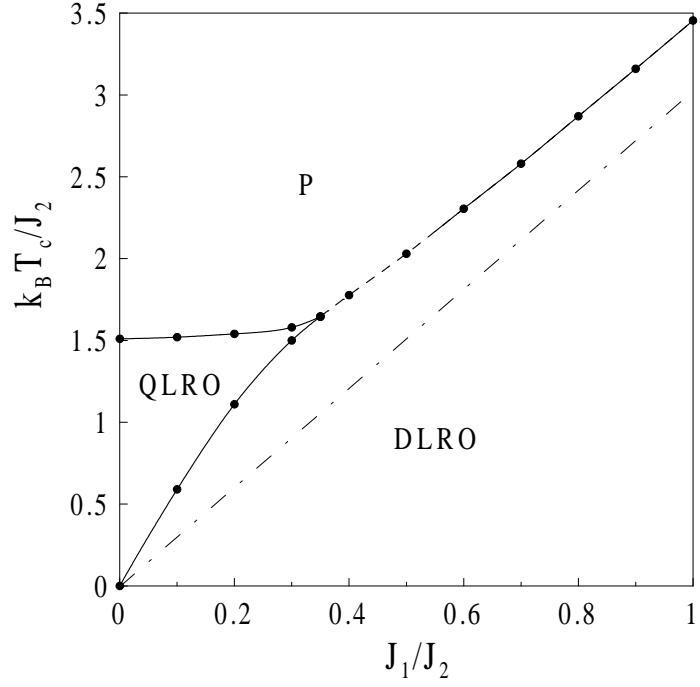}
\caption{Phase diagram in $(J_{1}/J_{2},k_{B}T_{c}/J_{2})$ space. The solid and dashed
lines correspond to second- and first-order transitions, respectively, and the dash-dot straight line
represents the boundary between paramagnetic and ordered regions in $(J_{1},k_{B}T_{c})$ space when the
biquadratic exchange is absent.}
\end{figure}

\newpage
\begin{table}
\caption{Critical indices $\nu_{Q},\gamma_{Q}$; $\nu_{M},\gamma_{M}$; and $\nu,\gamma$ for quadrupole,
dipole and simultaneous ordering, respectively.} \label{tab.1}
\begin{center}
\begin{tabular}{|c||c|c|c|}                                                                \hline
 $J_{1}/J_{2}$      & $\nu_{Q}$   & $\gamma_{Q}$   & $\gamma_{Q}$ (Ref.\cite{chen-etal2})  \\ \hline
 0                  & 0.667 $\pm$ 0.008    & 1.333 $\pm$ 0.018      & 1.32 $\pm$ 0.03                     \\ \hline
 0.2                & 0.660 $\pm$ 0.008    & 1.305 $\pm$ 0.011      & 1.30 $\pm$ 0.03                     \\ \hline \hline
                    & $\nu_{M}$   & $\gamma_{M}$   &                             \\ \hline
 0.1                & 0.642 $\pm$ 0.006    & 1.241 $\pm$ 0.011      & -                           \\ \hline
 0.3                & 0.654 $\pm$ 0.009    & 1.216 $\pm$ 0.006       & -                           \\ \hline \hline
                    & $\nu$       & $\gamma$       &                             \\ \hline
 0.8                & 0.677 $\pm$ 0.008    & 1.177 $\pm$ 0.015      & -                           \\ \hline

\end{tabular}
\end{center}
\end{table}


\newpage

\begin{thebibliography}{}

\bibitem{sivar72} J. Sivardiere,
        {\em Phys. Rev. B} {\bf 6}, 4284, (1972).
\bibitem{sivar73} J. Sivardiere, A.N. Berker and M. Wortis,
        {\em Phys. Rev. B} {\bf 7}, 343, (1973).
\bibitem{chen-levy1} H.H. Chen and P. Levy,
        {\em Phys. Rev. Lett.} {\bf 27}, 1383, (1971); {\em Phys. Rev. B} {\bf 7}, 4267, (1973).
\bibitem{chen-levy2} H.H. Chen and P. Levy,
        {\em Phys. Rev. B} {\bf 7}, 4284 , (1973).
\bibitem{micnas} R. Micnas,
        {\em J. Phys. C: Solid St. Phys.} {\bf 9}, 3307, (1976).
\bibitem{chaddha99} G.S. Chaddha and A. Sharma,
        {\em J. Magn. Magn. Mater.} {\bf 191}, 373, (1999).
\bibitem{chen-etal1} K.G. Chen, H.H. Chen, C.S. Hsue and F.Y. Wu,
        {\em Physica} {\bf 87A}, 629, (1977).
\bibitem{chen-etal2} K.G. Chen, H.H. Chen and C.S. Hsue,
        {\em Physica} {\bf 93A}, 526, (1978).
\bibitem{carmesin} H.-O. Carmesin,
        {\em Phys. Lett. A} {\bf 125}, 294, (1987).
\bibitem{tanaka} A. Tanaka and T. Idogaki,
        {\em J. Phys. Soc. Japan} {\bf 67}, 604, (1998).
\bibitem{campbell} M. Campbell and L. Chayes,
        {\em J. Phys. A} {\bf 32}, 8881, (1999).
\bibitem{nagata} H. Nagata, M.\v{Z}ukovi\v{c} and T.Idogaki - unpublished.
\bibitem{ferr-swen1} A.M. Ferrenberg and R.H. Swendsen,
        {\em Phys. Rev. Lett.} {\bf 61}, 2635 , (1988).
\bibitem{ferr-swen2} A.M. Ferrenberg and R.H. Swendsen,
        {\em Phys. Rev. Lett.} {\bf 63}, 1195 , (1989).
\bibitem{guillou} J.C. Le Guillou and J. Zinn-Justin,
        {\em Phys. Rev. Lett.} {\bf 39}, 95 , (1977).
\bibitem{janke} W. Janke,
        {\em Phys. Lett.} {\bf 148}, 306 , (1990).
\bibitem{allan} G.A.T. Allan and D.D. Betts,
        {\em Proc. Phys. Soc.} {\bf 91}, 341 , (1967).
\bibitem{vollmayr} K. Vollmayr, J.D. Reger, M. Scheucher and K. Binder,
        {\em Z. Phys. B} {\bf 91}, 113 , (1993).
\bibitem{ferr-landau} A.M. Ferrenberg and D.P. Landau,
        {\em Phys. Rev. B} {\bf 44}, 5081, (1991).

\end{thebibliography}
\end{document}